\documentclass[aps,pra,twocolumn,superscriptaddress,showkeys]{revtex4-1}

\usepackage{graphicx}
\usepackage{amsthm,amssymb,amsmath}
\usepackage{textcomp}
\usepackage{color}
\usepackage{braket}
\usepackage{gensymb}
\usepackage[euler]{textgreek}

\begin{document}

\title{Photorefractive effect in LiNbO$_3$-based integrated-optical circuits for continuous variable experiments}

\author{Fran{\c{c}}ois Mondain}
\address{Universit\'e C\^ote d'Azur, CNRS, Institut de Physique de Nice (INPHYNI), UMR 7010, Parc Valrose, 06108 Nice Cedex 2, France.}
\author{Floriane Brunel}
\address{Universit\'e C\^ote d'Azur, CNRS, Institut de Physique de Nice (INPHYNI), UMR 7010, Parc Valrose, 06108 Nice Cedex 2, France.}
\author{Xin Hua}
\address{Universit\'e C\^ote d'Azur, CNRS, Institut de Physique de Nice (INPHYNI), UMR 7010, Parc Valrose, 06108 Nice Cedex 2, France.}
\author{\'Elie Gouzien}
\address{Universit\'e C\^ote d'Azur, CNRS, Institut de Physique de Nice (INPHYNI), UMR 7010, Parc Valrose, 06108 Nice Cedex 2, France.}
\author{Alessandro Zavatta}
\address{Istituto Nazionale di Ottica (INO-CNR) Largo Enrico Fermi 6, 50125 Firenze, Italy}
\address{LENS and Department of Physics, Universit\`a di Firenze, 50019 Sesto Fiorentino, Firenze, Italy}
\author{Tommaso Lunghi}
\address{Universit\'e C\^ote d'Azur, CNRS, Institut de Physique de Nice (INPHYNI), UMR 7010, Parc Valrose, 06108 Nice Cedex 2, France.}
\author{Florent Doutre}
\address{Universit\'e C\^ote d'Azur, CNRS, Institut de Physique de Nice (INPHYNI), UMR 7010, Parc Valrose, 06108 Nice Cedex 2, France.}
\author{Marc P. De Micheli}
\address{Universit\'e C\^ote d'Azur, CNRS, Institut de Physique de Nice (INPHYNI), UMR 7010, Parc Valrose, 06108 Nice Cedex 2, France.}
\author{S\'ebastien Tanzilli}%
\address{Universit\'e C\^ote d'Azur, CNRS, Institut de Physique de Nice (INPHYNI), UMR 7010, Parc Valrose, 06108 Nice Cedex 2, France.}
\author{Virginia D'Auria}
\email{Virginia.DAuria@univ-cotedazur.fr}
\address{Universit\'e C\^ote d'Azur, CNRS, Institut de Physique de Nice (INPHYNI), UMR 7010, Parc Valrose, 06108 Nice Cedex 2, France.}

\date{\today}

\begin{abstract}
We investigate the impact of photorefractive effect on lithium niobate integrated quantum photonic circuits dedicated to continuous variable on-chip experiments.
The circuit main building blocks, \textit{i.e.} cavities, directional couplers, and periodically poled nonlinear waveguides are studied.
This work demonstrates that, even when the effect of photorefractivity is weaker than spatial mode hopping, they might compromise the success of on-chip quantum photonics experiments.
We describe in detail the characterization methods leading to the identification of this possible issue.
We also study to which extent device heating represents a viable solution to counter this effect.
We focus on photorefractive effect induced by light at 775\,nm, in the context of the generation of non-classical light at 1550\,nm telecom wavelength.
\end{abstract}


\maketitle


\section{Introduction}

With the development of quantum information technologies, photonic circuits are attracting an increasing interest as platforms for future out-of-laboratory realizations. Strong light confinement in waveguides guarantees efficient quantum state generation and manipulation in miniaturized structures which enables a progressive increase in architecture complexity thanks to high system stability and scalability~\cite{Review,Tanzilli2011}.\\
Lithium niobate (LiNbO$_3$) waveguides are particularly well adapted to quantum photonics, featuring easy in- and out-coupling to single-mode fibers, high nonlinear frequency conversion efficiencies, \mbox{electro-optical} properties, and low propagation losses~\cite{Alibart2016}. Largely used in single photon regime, LiNbO$_3$ chips are now moving also toward applications to continuous variable (CV) quantum optics, where quantum information is coded on light continuous spectrum observables.
CV quantum states produced at telecom wavelengths (around 1550\,nm) for quantum communication and computing~\cite{Leuchs2010} can be deterministically produced in LiNbO$_3$ by spontaneous parametric down conversion (SPDC) of near-infrared light (775\,nm)~\cite{Kaiser2016} and unambiguously discriminated by \mbox{on-chip} homodyne detection~\cite{Lenzini2018, Mondain2018}.
In comparison to single photon regime, CV experiments require higher pump powers at the input of the downconverters.
In such a condition, LiNbO$_3$ is likely to be affected by photorefractive effect: impurities and defect centers in the crystal cause a small absorption of the light traveling along the waveguide.
The photogenerated charges move out of the illuminated area and get trapped at the edges of the waveguide.
The resulting space-charge field leads to a refractive index modification that changes the spatial beam profile of the guided mode~\cite{Kostritskii2009}.
This can cause, in multimode waveguides, a coupling between the spatial modes (mode hopping)~\cite{Pal2015, Gunter2006}.
Metal doping (magnesium, zinc, indium, or hafnium) can be employed to mitigate photorefractivity but the fabrication techniques are not yet mature enough for being used in complex circuits~\cite{Volk94,Kong2012}. Photorefractive effect is partially or completely suppressed by operating undoped congruent LiNbO$_3$ at high working temperatures~\cite{Rams2000}, so as to increase charges' mobility.
However, higher temperature becomes rapidly unpractical and a compromise must be found between pump power and device temperature.\\
This paper bridges material science and quantum photonics aspects.
It investigates LiNbO$_3$ photonic circuits in working conditions that, although below the appearance of mode hopping, are submitted to photorefractive effect.
We study the impact of photorefractivity on the main building blocks of CV experiments in situations where the induced alteration is weak and by performing dedicated tests on the components. 
We focus on the on-chip generation and detection of squeezed light, \textit{i.e.} of optical quantum states exhibiting a noise level below the classical limit on one of their CV observables~\cite{Andersen2016}. Squeezed states lie at the very heart of many CV quantum-information protocols and are crucial for the heralded generation of non-Gaussian optical states as required for universal quantum computing~\cite{Leuchs2010}.
At the same time, they are very sensitive to losses as well as to generation or detection imperfections, hence they represent an extremely pertinent case-study for the analysis of unwanted effects occurring in integrated photonic systems. 
In the following, we illustrate the consequences of photorefractive effect on crucial elements of squeezing experiments, such as integrated cavities, directional couplers, and PPLN sources, components having already been successfully implemented on LiNbO$_3$ CV complex photonic circuits~\cite{Lenzini2018, Mondain2018, Stefszky2017}.
We will highlight, as a function of the pump power level and chip temperature, the careful optimization necessary to avoid a degradation of the squeezing. 

\section{Sample fabrication and characterization setup}

Ion-diffusion currently represents the most reliable technique to fabricate photonic integrated circuits on LiNbO$_3$~\cite{bazzan2015}. Depending on the specific fabrication procedure, the sensitivity to photorefractive effect varies.
Proton-exchanged (PE) waveguides exhibit the highest robustness to photorefractivity but suffer from a strong degradation of electro-optic and nonlinear optic properties that make them not suitable for integrating complex circuits.
Annealed proton-exchanged (APE) and reverse proton-exchanged (RPE) waveguides allow to partially recover the material properties at the price of higher sensitivity to photorefractive effect.
Eventually, titanium indiffused waveguides benefit from non-degraded LiNbO$_3$ properties but are strongly affected by photorefractivity, because of lower dark conductivity~\cite{Pal2015, Fujiwara1993}.\\
In this context, soft proton exchange (SPE) technique is particularly convenient for quantum photonics, both in single photon~\cite{Alibart2016} and CV regimes~\cite{Mondain2018}: it prevents degradation of the material properties during the manufacturing process leading to nonlinear efficiency better than those of  APE structures. In addition, it allows higher refractive index jumps providing a more efficient optical confinement with respect to RPE~\cite{Alibart2016}.

In this work, we will explore photorefractive effects in SPE waveguides. In this regard we note that photorefractivity of SPE structures is close to the one observed in APE and RPE waveguides~\cite{Caballero-Calero2006}. More in general, we stress that, although the specific numerical results are valid only for SPE waveguides, characterization strategies and considerations developed here can be generalized to devices fabricated via \textit{any} ion-diffusion technique.\\
In our analysis, we use in-house photonic integrated circuits, fabricated by means of 72 hours SPE at 300$\degree$C in benzoic acid buffered with lithium benzoate.
More details about the fabrication technique can be found in~\cite{Alibart2016, Chanvillard2000}.
Typical transmission losses yield 0.1\,dB/cm (1\,dB/cm) at 1550\,nm (775\,nm).
All the studied channel waveguides are 6\,\textmu m-wide, single-mode at telecom wavelengths, and slightly multimode at 775\,nm, where photorefractivity can therefore cause \mbox{mode hopping}.

\begin{figure}[ht]
  \centering
  \includegraphics[width=\linewidth]{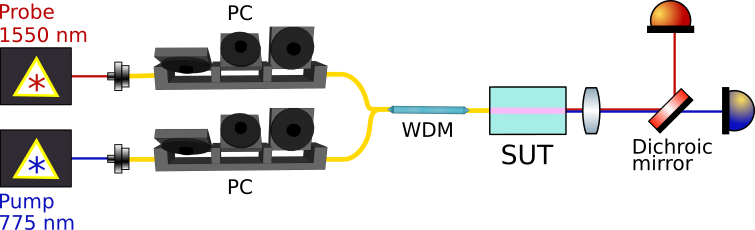}
	\caption{Schematics of the characterization setup used to investigate photorefractivity in LiNbO$_3$ quantum photonic circuits.
The probe beam is delivered by a EXFO T100S-HP tunable laser; the pump by a M-Squared SolsTiS laser.
PC: fiber polarization controllers.
WDM: fiber wavelength division multiplexer.
SUT: sample under test, \textit{i.e.} integrated cavity, directional coupler or periodically poled nonlinear waveguide.
Light sensors are Coherent OP-2 Vis and OP-2 IR.
Residual pump power on the probe beam detector is avoided thanks to the dichroic mirror (DMLP950T-Thorlabs), whose transmission at 775\,nm is less than 0.1$\%$.
Isolators after the probe and the pump lasers are not represented on the scheme.}
	\label{fig:setup}
\end{figure}

The experimental setup developed for the investigation of photorefractive effect is presented in Fig.\,\ref{fig:setup}: it employs a narrowband tunable laser at 1550\,nm as a probe and a tunable laser at 775\,nm as a pump inducing the photorefractive \mbox{space-charge field}.
Probe and pump beams are multiplexed through the same optical fiber and coupled to the sample under test (SUT).
Both optical sources are operated in continuous wave regime (CW) and have their polarization independently controlled.
At the output of the SUT, light is collected by a lens, its spectral components separated using a dichroic mirror, and focused onto suitable photodetectors.
The temperature of the system is stabilized by an active loop with an error of $\pm$5\,mK. We note that in our working conditions, 
photothermal drift only induces changes of $\sim 10^{-7}$\,/K in the refractive index~\cite{Birra2005}; this effect can be neglected compared to the one induced by photorefractivity, that is approximately three order of magnitude more important (see below).

\section{Cavity effect}

\subsection{Waveguide cavity made of chip's end-facets}
To date, the highest levels of squeezing have been obtained thanks to SPDC in resonant systems such as optical parametric oscillators~\cite{Schnabel2016}.
Integrated optics opens the possibility of fabricating monolithic squeezers that are more compact and stable than their bulk optics analogues, working either in single-pass or resonators.
This concept has been recently validated by the realisation  PPLN waveguides on ZnO-doped substrate that was used to demonstrate single-mode squeezing at 1550\,nm, with a noise reduction of about -6\,dB in single pass configuration~\cite{Furusawa2020}. At the same wavelengths, squeezing of -5\,dB has been measured at the output of a Ti-indiffused PPLN waveguide resonator~\cite{Stefszky2017} pumped with 30\,mW of CW powers at 775\,nm.
The observation of stronger squeezing values in such a resonant device was prevented by the occurence of photorefractive effect at higher SPDC pump powers.
\\
Beside instabilities caused by the appearance of mode hopping of the pump beam, photorefractive effect can indeed impact cavity based devices even at weaker power conditions.
More specifically, the waveguide refractive index variation, $\Delta n(t)$, induced by photorefractivity can result in a phase shift and, possibly, in a detuning from perfect cavity resonance that can degrade the squeezing level~\cite{Fabre1990}.
This effect can also affect efficient squeezing production in LiNbO$_3$ in single-pass configurations~\cite{Furusawa2007, Fejer2002, Mondain2018, Kaiser2016} where parasitic cavity effects can arise from Fresnel's reflections at the end-facets of the waveguide;
here, we focus on the effects of these parasite cavities. To his end, we take as SUT a 15\,mm long SPE waveguide with \mbox{end-facets} polished perpendicular to light propagation direction and uncoated.
The propagating modes indices are $n_{eff}$(@1550\,nm)=2.13 and $n_{eff}$(@775\,nm)=2.18.
Corresponding Fresnel's coefficients at the waveguide interfaces are $R_{wg \rightarrow air}\approx 0.14$ ($0.13$) per facet at 1550\,nm (775\,nm), and the device acts as a parasitic Fabry-Perot interferometer (FPI) with a low finesse $\mathcal{F}$=1.30 and a FWHM of 20\,pm~\cite{Regener1985}. The cavity enhancement to squeezing generation due to Fresnel's reflections at the sample facets is incremental. At the same time, a photorefractive-induced detuning from cavity resonance can result in an overall degradation of the device transmission and give rise to instabilities.

\subsection{Phenomenological description}

\begin{figure}[!h]
  \centering
  \includegraphics[width=\linewidth]{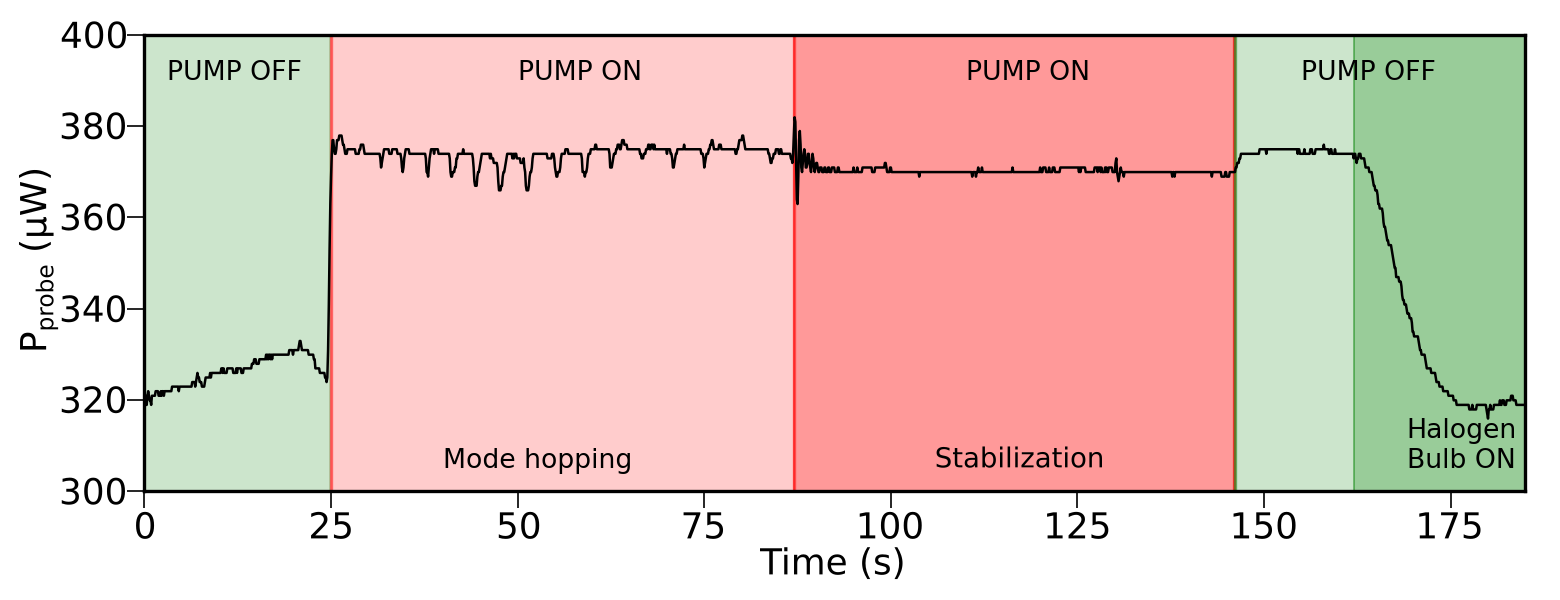}
	\caption{Transmitted power of a 1550\,nm wavelength probe signal as a function of time, measured in a channel waveguide with normal polished end-facets.
Red regions indicate when a pump beam at 775\,nm is injected in the waveguide.
Green regions indicate when the pump beam is off.
Between 25\,s and 85\,s, pump mode hopping affects the transmission of the probe beam. The power drop at 20\,s is due to unwanted mechanical misalignment.\label{fig:ON-OFF}}
\end{figure}

The complexity of the dynamics of the system is well illustrated in Fig.\,\ref{fig:ON-OFF} that shows the transmission trace of the probe at 1550\,nm, when the SUT working temperature is 30$\degree$C.
Here, green regions indicate the time-slots in which only the probe beam is sent to the SUT while red areas refer to the case in which the system is fed also with a pump beam of 5\,mW power.
In such weak pumping regime, we already observe photorefractive effect.
The pump irradiation changes the effective refractive index seen by the probe at 1550\,nm and, in turn, its transmitted power through the FPI~\cite{Hellwig2011}.
This is clearly visible between 25\,s and 85\,s when regular dips appear in the transmission trace, due to refractive index variation associated with mode hopping of the pump beam.
These mode fluctuations spontaneously reduce until the system stabilizes on a given mode as witnessed by the probe transmission after 85\,s.
At 140\,s the pump laser is turned off with little impact on the space-charge field.
This is related to the long lifetime of the trapped electrons.
At 165\,s we illuminate the sample with a high-intensity halogen bulb (1.4 W) to accelerate the redistribution rate of the trapped electrons so as to restore the initial material conditions and beam transmission. 

\subsection{Model of the waveguide cavity}
\begin{figure*}[!htb]
\centering
\begin{tabular}{c c}
  \includegraphics[width=0.45\linewidth]{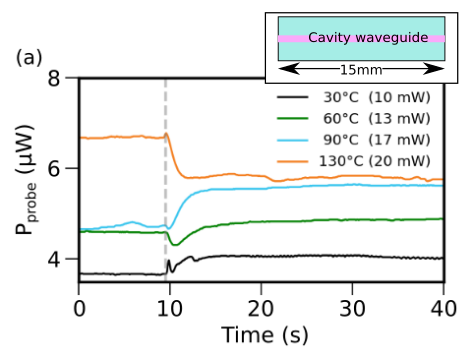} & \includegraphics[width=0.45\linewidth]{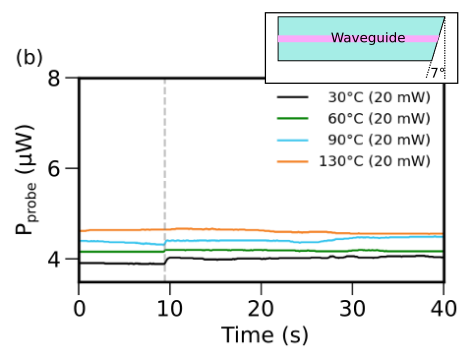}
\end{tabular}
\caption{Transmitted power at 1550\,nm versus time measured in a channel waveguide in a resonant (a), and non-resonant (b) configuration.
Grey-dashed line indicates the onset of the pump beam. Insets show a simple schematics of the chip.\label{fig:Temp-Pprobe}}
\end{figure*}

The large variety of phenomena shown in Fig.\,\ref{fig:ON-OFF} suggests why photorefractivity is generally hard to predict and control. The usual countermeasure consists in increasing the SUT temperature around 100$\degree$C.
However, this only shifts the occuring of photorefractive effect towards higher pump powers, without completely suppressing it.
We illustrate this behavior in Fig.\,\ref{fig:Temp-Pprobe}a, where we show the transmission traces for the probe beam at different heating temperatures with a pump injection starting at t=10\,s.
At low temperatures, the pump power was adjusted so as to avoid mode hopping. 
For all cases, the final transmission for stabilized waveguide is different from the original one. Before reaching this point, a transient oscillatory behaviour might be observed.

To better understand the interplay between cavity and photorefractive effect, we recall the transmission of the Fabry-Perot waveguide resonator~\cite{hecht2017}:
\begin{equation}
T(t)= \frac{1}{1+\mathcal{F}\sin^2\left( \frac{2\pi L\cdot(n_{eff}+\Delta n(t))}{\lambda }\right)},
\end{equation}

\noindent with $\mathcal{F}=\frac{4R}{(1-R)^2}$ is the finesse of the cavity, $\lambda$ the probe wavelength, $L$=15\,mm the length of the chip, $n_{eff}$ the effective index of the mode without photorefraction, and $\Delta n(t)$ the pump-induced photorefractive index modification. Since photorefractive effect barely changes Fresnel's coefficients, we can consider $\mathcal{F}$ constant.\\

From the $sin^2$ term, we see that for weak variation of $\Delta n$ over time, the transmitted power can show an oscillatory behaviour, half a period corresponding to an index variation $\Delta n\sim 2.6\times 10^{-5}$.
This kind of variation can be observed in the $T=30\degree$\/C curve of Fig.\,\ref{fig:Temp-Pprobe}a between 10 and 15\,s, giving an overall index change $\Delta n\sim 8\times 10^{-5}$; for higher temperatures, $\Delta n$ is weaker and oscillations disappear. We note that transmission is affected by photorefractivity even at 130$\degree$C. At such high temperature, a small, time-independent, refractive index shift can be observed~\cite{Amodei1971}. This constant shift slightly modifies the cavity optical resonance condition and, as a consequence, the transmitted power at the probe wavelength. 
Beside the use of anti-reflection coatings, the most pertinent way to suppress parasitic FPI is to polish the chip end-facets with an angle in order to prevent guiding of light reflected at the interface. We confirm this concept by employing as a SUT a chip with output facets polished with a 7$\degree$ angle~\cite{Takaya2003}.
The corresponding transmissions are shown in Fig.\,\ref{fig:Temp-Pprobe}b: all measurements are taken at a maximum pump power of 20\,mW.
The signature of photorefraction is removed or negligible for all working temperatures.
Residual effects are due to the photorefractive inducted $\Delta n$, that changes the fiber-to-waveguide coupling at the input of the non-resonant device, thus affecting the measured  transmitted power.

\begin{figure}[!htb]
\centering
\begin{tabular}{c}
 \includegraphics[width=0.8\linewidth]{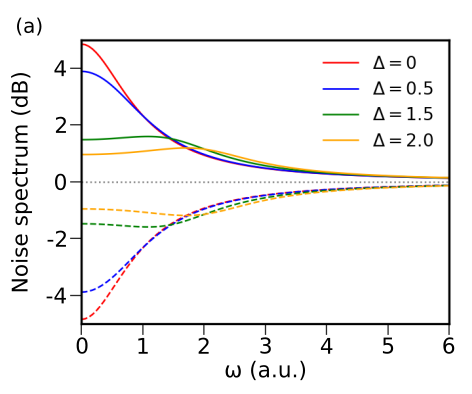}\\ \includegraphics[width=0.8\linewidth]{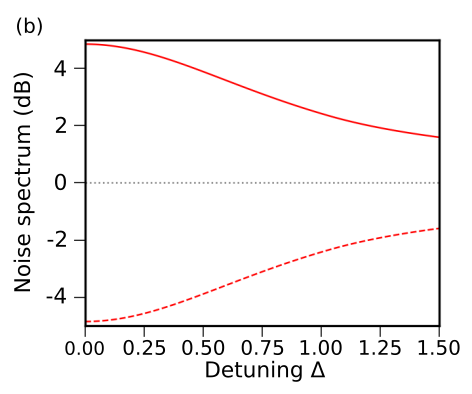}
\end{tabular}
 \caption{(a) Squeezing (dashed) and anti-squeezing (solid line) spectra emitted by SPDC in a resonant nonlinear waveguide cavity and (b) corresponding optimal squeezing and anti-squeezing as functions of the detuning.
In the simulations, $\Delta$ represents the detuning normalized to the cavity linewidth. We assume no active feedback system ensuring the cavity resonance condition.\label{fig:SQ}}
 \end{figure}

Eventually, we note that, in CV quantum optics, beside modifying the sample transmission, the detuning from perfect cavity resonance (\textDelta) induced by \textDelta n(t) can affect the dynamics of squeezing generation~\cite{Fabre1990}.
For example, Fig.\,\ref{fig:SQ} a shows the quantum noise spectrum of single-mode squeezed light at the output of a resonant PPLN waveguide~\cite{Stefszky2017}.
The detuning is normalized to the cavity linewidth.
The squeezing spectrum (dashed lines) exhibits a minimum at a frequency that depends on the detuning and shows a noise reduction less important than the one expected in the perfectly resonant case ($\Delta=0$).
This effect is more obvious when \textDelta ~increases. This is represented in Fig.\,\ref{fig:SQ}b where the best squeezing (antisqueezing) is represented as functions of the detuning.
As for a numerical example, for a cavity of 15\,mm, with mirror reflectivity of 0.77 and 0.99 as in~\cite{Stefszky2017}, a variation of $\Delta n=2.6\times10^{-5}$ would correspond to a normalized detuning of $\Delta=$1.5 and lead to a degradation of squeezing, from an initial value of -5\,dB, to only -2\,dB. Similar calculations show that the same detuning would reduce an initial squeezing of -10\,dB to only $\approx$-5\,dB.

\section{Directional couplers and homodyne detection}

\begin{figure}[!htb]
  \centering
  \includegraphics[width=\linewidth]{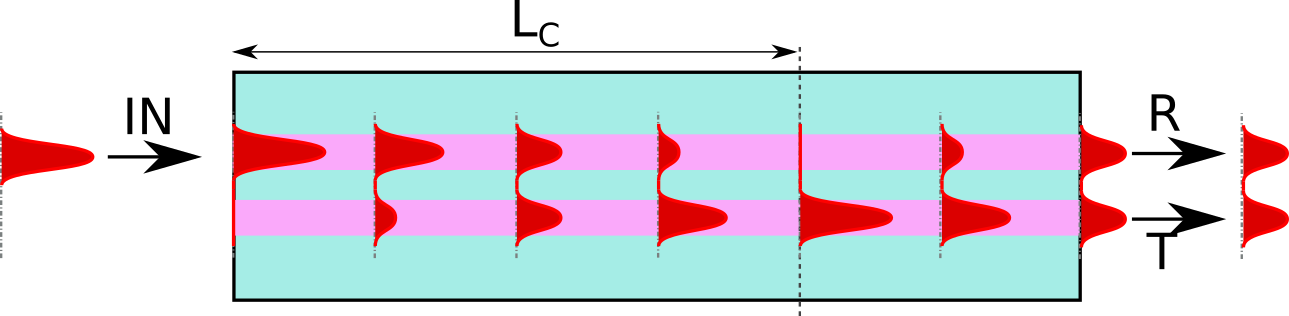}
	\caption{Schematics of evanescent-wave coupling balanced beam splitter, here with a total length of 3/2\,L$_c$.
	The two output ports are $R$ reflected (\textit{i.e.} the same in which injection is made), and $T$ transmitted. $L_C$ stands for coupling length, \textit{i.e.} the propagation length after which all the optical power is in the transmitted arm.\label{fig:BS}}
\end{figure}

Directional couplers are basic building blocks for any photonic integrated circuit as they permit to design light routing, (de)multiplexing and interference stages.
We consider here devices exploiting evanescent-wave coupling to obtain a periodic transfer of light intensity from a waveguide to another: when two single-mode waveguides are close enough, their evanescent fields spatially overlap and give rise to a coupling between the two structures, as sketched in Fig.\,\ref{fig:BS}.

The fraction of optical power transferred to the adjacent waveguide is generally said to be \textit{transmitted}, while the rest is \textit{reflected}; by extension we will call the two waveguides reflecting ($R$) and transmitting ($T$), with propagation constants $\beta_R = \frac{2\pi n_R}{\lambda}$ and $\beta_T = \frac{2\pi n_T}{\lambda}$, respectively.
If $\Delta\beta=0$, light will periodically undergo a perfect transfer from a waveguide to the other during its propagation.
All optical power is found in the transmitting arm after a distance referred to as the coupling length, $L_c = \frac{\pi}{2k}$ where $k$ is the coupling constant that depends on the effective refractive indices of the modes and on the distance separating the waveguides.

In the context of CV integrated circuits, directional couplers are needed for multiple applications.
On the one hand, strongly unbalanced couplers are used to pick up a fraction of a non-classical state and send it to a single photon detector for the heralded preparation of non-Gaussian states~\cite{Huang2015}.
On the other hand, perfectly balanced couplers are mandatory for homodyne detection, in which the properties of a state under scrutiny are retrieved by making it interfere with a local oscillator~\cite{Mondain2018}.
Refractive index change induced by photorefractivity can affect the symmetry of directional couplers ($\Delta \beta =0$) and shift their working point. This effect is detrimental in most of practical situations.

\begin{figure*}[!htb]
\centering
\begin{tabular}{l r}
\includegraphics[width=0.35\linewidth]{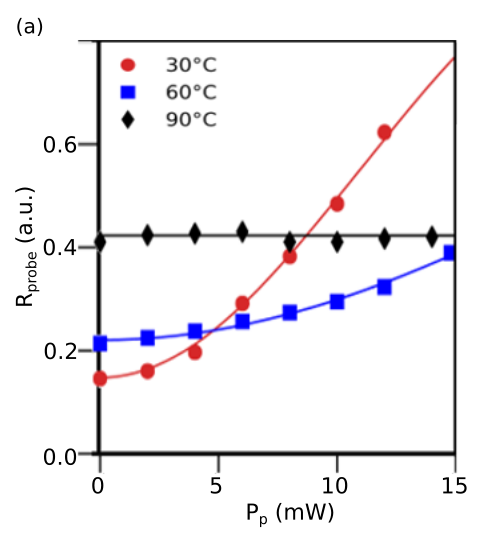} & \includegraphics[width=0.35\linewidth]{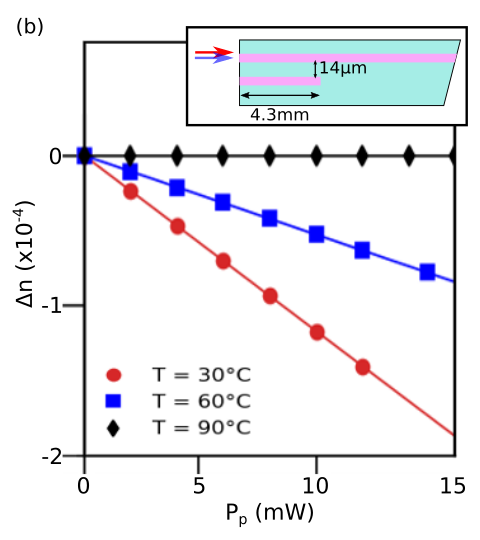}
\end{tabular}
\caption{(a)~Coupler reflectivity, $\mathrm{R}$, at 1550\,nm as a function of the pump power coupled in the waveguide at 30$\degree$C (red dots), 60$\degree$C (blue squares), 90$\degree$C (black diamonds); solid lines represent the fits.
(b)~Refractive index variation of the integrated coupler, $\Delta n$, at 1550\,nm as a function of the pump power.
Inset: coupler chip schematics, the two waveguides are separated by 14\,\textmu m (center-to-center), the coupling length $L_c$ at 1550\,nm is $3.43$\,mm at 30$\degree$C, $3.27$\,mm at 60$\degree$C and $2.96$\,mm at 90$\degree$C.\label{fig:Temp-coupler}}
\end{figure*}

We investigate the influence of photorefractivity thanks to a passive directional coupler (sketched in the inset of Fig.\,\ref{fig:Temp-coupler}a) consisting in a 15\,mm long chip with its output facet angle-polished to prevent parasitic FPI.
It comprises two waveguides of 6\,\textmu m width with 14\,\textmu m separation, corresponding to a coupling k=0.46\,mm$^{-1}$, \textit{i.e.} $L_c=3.43$\,mm at $T=30\degree$C for $\lambda_{probe}=1550$\,nm. The coupler length is 4.3\,mm long.
At the pump wavelength, the coupling between the waveguides is negligible.
This geometry, although made of only one output (the reflection waveguide), is perfectly suitable to study the device sensitivity to photorefractivity.

As shown in Fig.\,\ref{fig:Temp-coupler}a, both pump and probe beams are injected in the coupler and the probe beam exiting the chip is analyzed.
When the pump beam is in the reflection waveguide, it modifies its refractive index $n_R \rightarrow n_R+\Delta n$, changing the propagation constant to $\Delta\beta = \frac{2\pi\Delta n}{\lambda}$.
Correspondingly, the fraction of reflected probe power is: 

\begin{widetext}
\begin{equation}\label{eqn:Rprobe}
\mathrm{R}_{probe}=\frac{P_R}{P_0}=1-\frac{4k^2}{4k^2+\Delta\beta^2}\cdot\mathrm{sin}^2\left(\frac{L\cdot\sqrt{4k^2+\Delta\beta^2}}{2}\right),
\end{equation}
\end{widetext}

\noindent with $P_{0}$ ($P_{R}$) the input (reflected) probe power.
Measured coupler reflectivity at 1550\,nm is plotted in Fig.\,\ref{fig:Temp-coupler}a as a function of the pump power $\mathrm{P}_\mathrm{p}$ at 775\,nm.
Experimental data have been fitted using \mbox{Eq.\,(\ref{eqn:Rprobe})} in order to infer $\Delta n$. We assume a dependence of $\Delta n$ upon the pump power of the form $\Delta n = -\frac{a\mathrm{P}_\mathrm{p}}{b+c\mathrm{P}_\mathrm{p}}$ where $a$, $b$ and $c$ are constants determined by the material~\cite{Ruske2003,Fujiwara1992}.
Obtained $\Delta n$ values are shown in Fig.\,\ref{fig:Temp-coupler}b. We note that, for studied pump powers, the dependence of $\Delta n$ with $\mathrm{P}_\mathrm{p}$ is linear and does not saturate. The experimental values are consistent with Fig.\,\ref{fig:Temp-Pprobe}b and compatible with results published in the literature~\cite{Ruske2003}. 
As for an example, a pump power of 10\,mW induces a $\Delta n\sim 10^{-4}$, close to the value of $8\cdot10^{-5}$ as found from the FPI analysis.
The difference between the two measured $\Delta n$ values can be explained by a combination of sample geometrical variations, due to the fabrication process, and measurement uncertainties (of the order of $\leq10\%$).
Note that, as expected, higher temperatures reduce the photorefractivity; above 90$\degree$\,C, photorefractivity impact on the coupler is negligible.\\

\begin{figure*}[!htb]
\centering
\begin{tabular}{l r}
\includegraphics[width=0.4\linewidth]{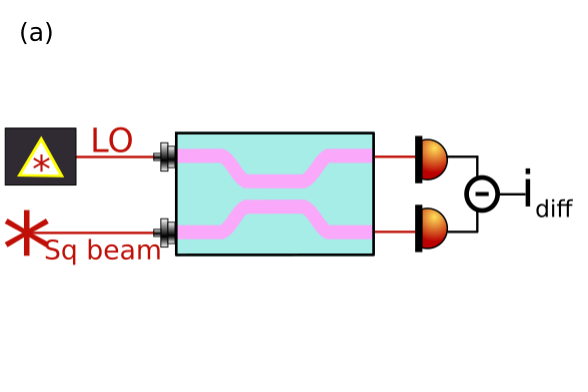} &  \includegraphics[width=0.4\linewidth]{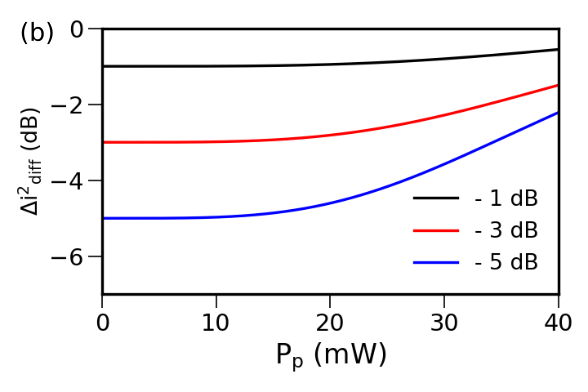}
\end{tabular}
\caption{(a)~Schematics of a generic integrated homodyne detection setup; LO: Local oscillator; Sq beam: squeezed beam; $i_{\mathrm{diff}}$: photo-current difference between the detectors.
(b)~Calculated maximal measurable squeezing $\Delta i^2_{\mathrm{diff}}$ when the splitting ratio of the homodyne coupler is altered by pump-induced photorefractivity for 3 different initial squeezing levels.
We assume the coupler working at 30$\degree$\/C.
We consider a perfect coherent local oscillator without classical noise.\label{fig:homodyne}}
\end{figure*}

We can apply these findings to the discussion on performances of an integrated homodyne detector as the one implemented in~\cite{Mondain2018}.
As sketched in Fig.\,\ref{fig:homodyne}a, in CV quantum optics, information on an optical beam is obtained by mixing it on a 50/50 coupler with a reference beam called local oscillator (LO) and then by detecting the intensities of the two outputs beams.
Single-mode squeezing level can be obtained from the noise variance of the photocurrent difference, $\Delta i^2_{\mathrm{diff}}(\mathrm{R})$.
For a generic coupler of reflectivity $\mathrm{R}$ and transmissivity $\mathrm{T} = 1-\mathrm{R}$, the measured quantum noise is:

\begin{widetext}
\begin{equation}\label{eqn:Deltai}
\Delta i^2_{\mathrm{diff}}(R)= \vert \alpha\vert ^2 \biggl[(\mathrm{T}-\mathrm{R})^2+4\mathrm{R}\cdot\mathrm{T}\cdot(e^{2s}\mathrm{sin}^2{\phi}+e^{-2s}\mathrm{cos}^2{\phi})\biggr] +\mathcal{O}(\vert \alpha\vert ^3),
\end{equation}
\end{widetext}

\noindent where $s$ is the squeezing parameter, $\vert \alpha\vert$ is the LO field amplitude, and $\phi$ is the relative phase between LO and the squeezed beam. In the previous expression, we consider a perfect coherent local oscillator without classical noise.
The noise of the squeezed observable (typically the electromagnetic field phase quadrature~\cite{Leuchs2010}) is $e^{-2s}$, while the anti-squeezing on its conjugate observable (the amplitude quadrature) is $e^{2s}$.
In the following we assume $\phi=0$ for simplicity and focus only on squeezing detection.
For a balanced coupler, \textit{i.e.} $\mathrm{T}=\mathrm{R}=1/2$, the terms from unbalanced LO disappear and residual classical noise contributions are fully suppressed. In this optimal condition, $\Delta i^2_{\mathrm{diff}}(1/2)=\vert \alpha \vert ^2 e^{-2s}$.
In experiments, to quantify the noise reduction with respect to a reference noise, squeezed light before the coupler is blocked so as to measure the vacuum noise (shot noise level) $\vert \alpha\vert ^2$ used to normalize $\Delta i^2_{\mathrm{diff}}$.
In integrated realizations, residual pump beam from the squeezing stage can circulate in the homodyne coupler and induce photorefractivity: this can change the splitting ratio either in the measurement or in the calibration step.
From the $\Delta n$ values of Fig.\,\ref{fig:Temp-coupler}b, we can evaluate the impact of photorefractivity on squeezing homodyne measurement. Fig.\,\ref{fig:homodyne}b shows the measured squeezing level as a function of the residual pump power circulating in the homodyne optical coupler.
The estimation is made for a coupler with 14\,\textmu m-waveguide separation, working at 30$\degree$\/C and designed to have a perfectly balanced splitting in absence of optical pumping at 775\,nm. Fig.\,\ref{fig:homodyne}b shows that the splitting ratio alteration induced by photorefractivity can significantly reduce the measured squeezing level; this underestimation can easily remain unnoticed without a preliminary proper characterization of the photorefractivity. 

Empirically, we find that a temperature of 90$\degree$\/C is sufficient to suppress photorefractivity when the pump power is below 15\,mW.
However, the total effect is strongly related to the exact configuration (coupling constant, coupling length, substrate, pump power).
To conclude this section we note that integrated wavelength division multiplexer (WDM) have already been integrated on LiNbO$_3$ photonic chips~\cite{Lenzini2018}.
Provided they are added before the homodyne detection stage, this kind of devices could be a solution to suppress residual pump power and, in turns, photorefractivity in the integrated coupler.
However, such a realization would come at the cost of an increased complexity and additional losses in the circuit.

\section{Spontaneous parametric down conversion}

We investigate the impact of photorefractivity on the nonlinear properties of a periodically poled lithium niobate waveguide (PPLN/w).
In particular, due to its role in squeezing generation, we discuss the evolution of spontaneous parametric down conversion (SPDC) at different pump regimes and temperatures. 
The SPDC spectrum is determined by energy conservation and phase matching conditions that rule the conversion of pump photons ($p$) at 775\,nm, into pairs of signal ($s$) and idler ($i$) photons at telecom wavelength ($\approx$1550\,nm).
These can be written, in a PPLN medium, as:

\begin{equation}
\label{eqn:QPM}
\frac{1}{\lambda_p}=\frac{1}{\lambda_i}+\frac{1}{\lambda_s}, \hspace{1.5 cm} \frac{n_p}{\lambda_p}-\frac{n_i}{\lambda_s}-\frac{n_s}{\lambda_s}=\frac{1}{2\Lambda},
\end{equation}

\noindent where $\lambda_j$ is the wavelength of the beam ($j=p,s,i$), $n_j$ the corresponding refractive index, and $\Lambda$ the poling period~\cite{Alibart2016}.
We note that the generation of single mode squeezing is achieved when signal and idler are fully degenerate~\cite{Kaiser2016}, in the general case two mode squeezing is obtained~\cite{Heidmann1987}.
As the quasi-phase matching condition depends on the material refractive indices, photorefractive effect can considerably modify the SPDC working point. 

\begin{figure}[!htb]
  \centering
  \includegraphics[width=\linewidth]{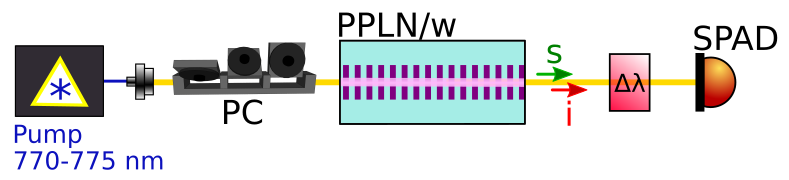}
  \caption{SPDC experiment schematics; Pump: tunable laser (Coherent Mira); PC: polarization controller; PPLN/w: in-house 15\,mm long PPLN soft-proton-exchanged waveguide; $\Delta\lambda$: tunable spectral bandpass filter (Yenista XTM-50); SPAD: InGaAs/InP single-photon avalanche diode (ID Quantique Id220), with efficiency set to 10\% and dead time to 15\,\textmu s; $s$: signal, $i$: idler.}
  \label{fig:setup-PPLN}
\end{figure}

The experimental setup used to investigate the impact of photorefractivity on SPDC is depicted in Fig.\,\ref{fig:setup-PPLN}.
A tunable pump laser delivering a wavelength of 770.73\,nm (774.63\,nm) is injected through a fiber into a in-house 15\,mm long PPLN SPE waveguide working at 30$\degree$\/C (90$\degree$\/C), designed to have its SPDC degeneracy around 1550\,nm.
The waveguide is polished so that incident beams are orthogonal to end-facets (0$\degree$-cut).
To comply with cavity effects, before acquiring the spectrum, we wait for the resonant system stabilization (see Fig.\ref{fig:Temp-Pprobe}).

\begin{figure*}[!htb]
\centering
\begin{tabular}{l r}
\includegraphics[width=0.45\linewidth]{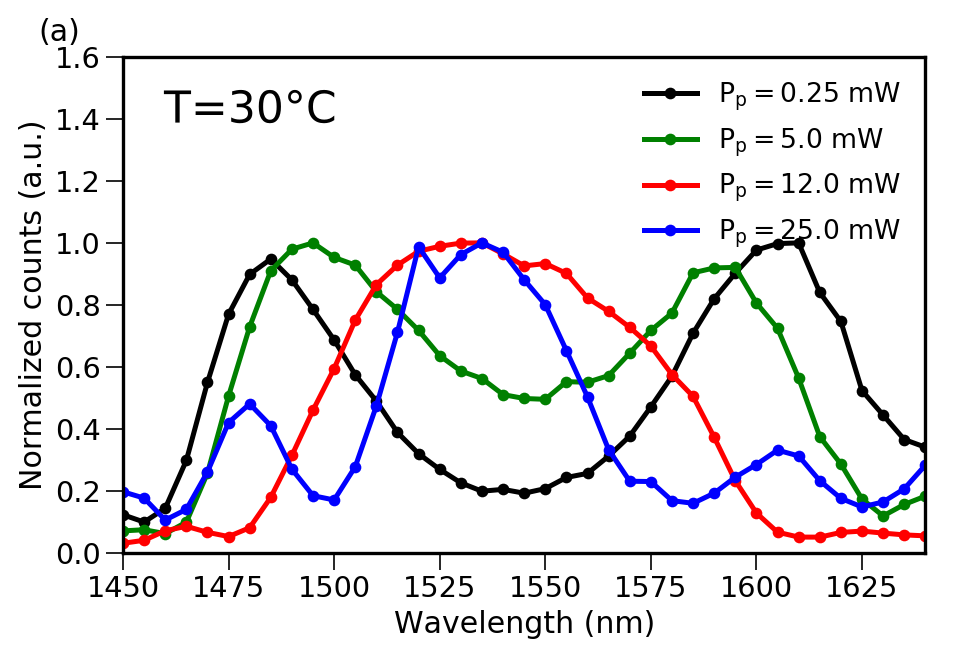} & \includegraphics[width=0.45\linewidth]{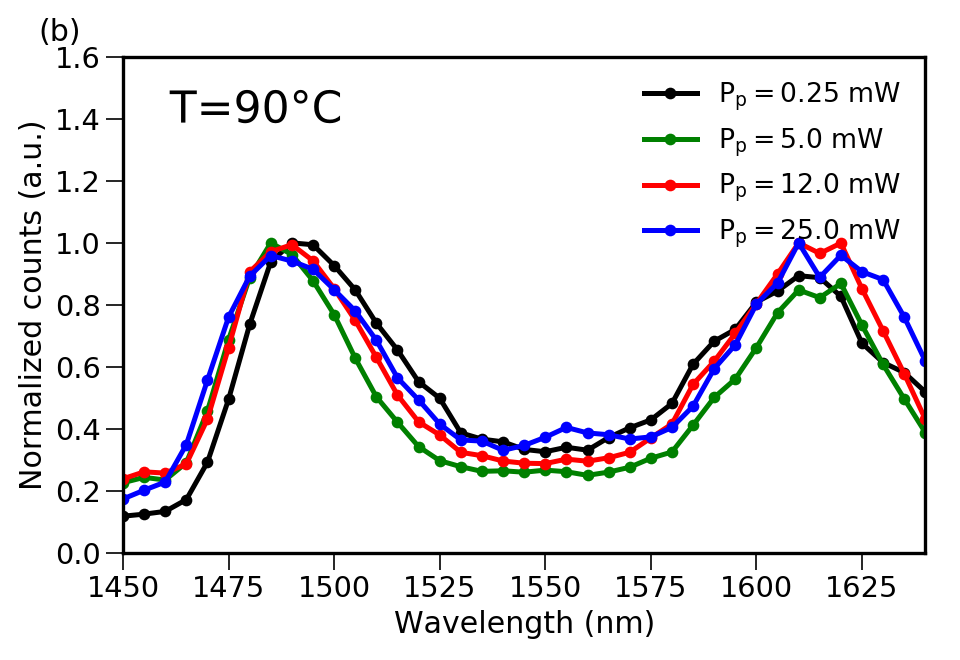}
\end{tabular}
\caption{SPDC normalized spectra for different pump powers coupled in the PPLN waveguide at (a) 30$\degree$\/C and (b) 90$\degree$\/C.\label{fig:SPDC}}
\end{figure*}

At the output of the sample, the generated photon pairs are coupled to an optical fiber, spectrally scanned by using a tunable filter, and measured by a single-photon detector.
This characterisation method is typical for the analysis of SPDC in quantum optics~\cite{Lunghi2012}.
The dark count rate is on the order of $\sim$1\,kcounts/s.
Note that, for all of the investigated powers, the sample works in a non-stimulated regime. As the photon pair emission rate is proportional to the pump power~\cite{Tanzilli2001}, an in-line fiber attenuator (not represented in Fig.\,\ref{fig:setup-PPLN}) is used downstream the sample, in order to keep the count rate below 20\,kHz to avoid detector saturation.
\\
The normalized resulting SPDC spectra for different pump powers ($\mathrm{P}_\mathrm{p}$) and temperatures are shown in Fig.\,\ref{fig:SPDC}a (30$\degree$\/C) and Fig.\,\ref{fig:SPDC}b (90$\degree$\/C).
The pump wavelength is adjusted for each temperature in order to be out of degeneracy when $\mathrm{P}_\mathrm{p}$=0.25\,mW.
At low temperature, as expected, the photorefractive effect due to the pump power, changes the phase-matching condition.
When $\mathrm{P}_\mathrm{p}$ increases, the emission reaches the degeneracy such that idler and signal emission spectra to overlap.
Similar results have been found in~\cite{Pal2015} for second-harmonic generation where a blue shift of the phase-matching wavelength has been observed.
We note that, as found for the directional coupler, phase matching is insensitive to the pump beam influence when the device is set to a temperature of 90$\degree$\/C.\\

\begin{figure}[!htb]
  \centering
  \includegraphics[width=0.95\linewidth]{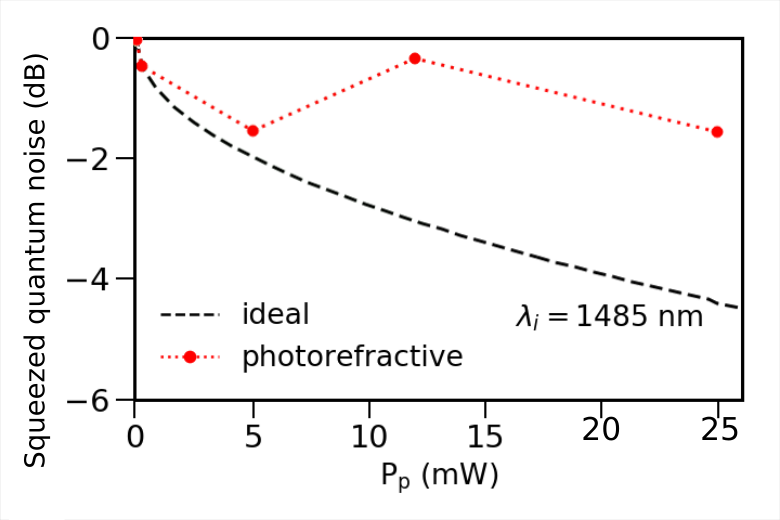}
  \caption{Squeezing for an ideal PPLN (dashed line) and for a PPLN that suffers from photorefraction (red circles), calculated using the SPDC efficiency spectra reported in Fig.\,\ref{fig:SPDC}\,a.\label{fig:idealSQ}}
\end{figure}

In this context, photorefractivity affects directly the squeezing generation. Indeed, the squeezing parameter, $s$ is equal to $\mathrm{\mu}\sqrt{\mathrm{P}_\mathrm{p}}$~\cite{Kaiser2016}, where $\mathrm{\mu}$ is the nonlinear efficiency, often considered as constant in PPLN waveguides.
However, we have seen that photorefractivity modifies the phase matching conditions, and $\mu$ then depends on $\mathrm{P}_\mathrm{p}$ via the refractive indices in Eq.\,(\ref{eqn:QPM}).\\
Let us compare an ideal PPNL waveguide, \textit{i.e.} not affected by photorefractivity (constant nonlinearity), to a realistic one (submitted to photorefractivity).
In the first case, we take $\mu$~=~0.101\,mW$^{-1/2}$ as reported in the literature~\cite{Kaiser2016}.
In the second case, we take into account the degradation of the quasi-phase matching due to photorefractivity from the SPDC spectra depicted in Fig.~\ref{fig:SPDC}a.
Both reachable squeezing levels are shown in Fig.\,\ref{fig:idealSQ}.
The shifted-QPM, photorefractive PPLN (red dotted line) is strongly degraded compared to the ideal, photorefraction-free PPLN (black dashed line).
Therefore, to optimize the squeezing level, proper and careful characterisations must thus be realized on the device working conditions in order to comply with photorefractivity and to compensate for the unwanted spectral shift, by fine adjustement of the temperature and pump wavelength. 

\section{Conclusions}

Integrated photonics enable a continuous increase of optical function complexity.
This is extremely advantageous for quantum information technology and in particular in CV quantum photonics where losses and stability need to be carefully controlled.
In this paper we have focused on LiNbO$_3$-based integrated photonic circuits which represent one among the most reliable platforms for CV-quantum regime and squeezing experiments.
However, LiNbO$_3$ notoriously suffers from photorefractivity and this is particularly crucial for squeezing as it requires strong pump beams at short wavelengths. 
Here, we have derived a framework to analyze the impact of photorefractive effect on integrated cavities, directional couplers, and nonlinear waveguides which are building blocks for any CV integrated photonic circuits.
We have demonstrated that photorefractivity might compromise the success of an experiment even in conditions far from mode hopping. A temperature increase is a viable solution to reduce photorefractivity.
For the specific case-studies investigated in this work, an operating temperature of at least 90$\degree$\/C is sufficient to suppress the impact of photorefractivity on waveguide coupling or SPDC process. 
Nevertheless, this strategy, even above 120$\degree$C, doesn't apply for resonators and cavities which are extremely sensitive to refractive-index variations.
Therefore, particular attention should be taken to avoid parasitic cavities.
If cavity is needed, the use of metal doped-LiNbO$_3$ becomes mandatory above a certain pump power that depends on the specific waveguide technology.
We believe this work might be a useful guideline to efficiently characterize LiNbO$_3$ integrated photonic chips for squeezing experiments.

\section*{Funding}
This work was conducted within the framework of the project OPTIMAL granted through the European Regional Development Fund (Fond Europeen de developpement
regional, FEDER). The authors acknowledge financial support from the Agence Nationale de la Recherche (HyLight ANR-17-CE30-0006-01, SPOCQ ANR-14-CE32-0019-03, Q@UCA ANR-15-IDEX-01).
\section*{Disclosures}
The authors declare no conflicts of interest.


\end{document}